# Considerations for Using the Vasculature as a Coordinate System to Map All the Cells in the Human Body

**Griffin M Weber[1], Yingnan Ju[2], Katy Börner[2]**

[1]Department of Biomedical Informatics, Harvard Medical School, Boston, MA, USA

[2]Department of Intelligent Systems Engineering, School of Informatics, Computing, and Engineering, Indiana University, Bloomington, IN, USA

\* Correspondence:
Katy Börner
katy@indiana.edu

**Abstract**

Several ongoing international efforts are developing methods of localizing single cells within organs or mapping the entire human body at the single cell level, including the Chan Zuckerberg Initiative's Human Cell Atlas (HCA), and the Knut and Allice Wallenberg Foundation's Human Protein Atlas (HPA), and the National Institutes of Health's Human BioMolecular Atlas Program (HuBMAP). Their goals are to understand cell specialization, interactions, spatial organization in their natural context, and ultimately the function of every cell within the body. In the same way that the Human Genome Project had to assemble sequence data from different people to construct a complete sequence, multiple centers around the world are collecting tissue specimens from a diverse population that varies in age, race, sex, and body size. A challenge will be combining these heterogeneous tissue samples into a 3D reference map that will enable multiscale, multidimensional Google Maps-like exploration of the human body. Key to making alignment of tissue samples work is identifying and using a coordinate system called a Common Coordinate Framework (CCF), which defines the positions, or "addresses", in a reference body, from whole organs down to functional tissue units and individual cells. In this perspective, we examine the concept of a CCF based on the vasculature and describe why it would be an attractive choice for mapping the human body.

## 1    Introduction

When examining coordinate systems that could be used for different biological entities, specific characteristics should be considered. For example, chromosome number and base pair position form a *linear* coordinate system that is sufficient to define location in the reference genome [Lander 2001]. However, a coordinate system for the entire human body, which can serve the needs of HuBMAP, HCA, HPA, and other mapping efforts, is far more complex for several reasons [HuBMAP Consortium 2019, Uhlén 2005, Rozenblatt-Rosen, Regev 2017]. First, cells live in three dimensions; and, additional information is needed to describe the size and rotational orientation of any tissue sample used to generate the data. Second, while genome base pair position does not depend on how DNA is folded and packed into the nucleus, the positions of individual elements (e.g., cells, organs) within the body are dynamic, as they may change with skeletal and muscle movements, gravity, breathing, heart beating, and other functions and forces that distort tissue. Third, although the variability of DNA between people is less than 1%, human body geometry spans a wide range of heights, weights, and shapes, which vary by sex and race and change significantly over a person's lifespan.

One option is to use a *3D Cartesian* coordinate system to describe position within the body. Cartesian coordinate systems have the benefit of being familiar and easily understood. With the body in standard anatomical position, left-right, caudal-cranial, and posterior-anterior form three perpendicular (x-y-z) axes. For smaller scale analyses, centering the origin on an anatomic landmark might be most useful and potentially less variable between different people. For example, it would be easier to measure the distance of a tissue sample to the superior pole of the kidney than to the top of a person's head. The disadvantage of Cartesian coordinates is that its axes do not follow the natural shape of the body. As a result, creating a reference body map by digitally "stitching" together tissue specimens collected from people of different sizes would require a complex process of warping, rotating, and aligning the data. Doing this in a way that scales to the entire body while maintaining cellular-level precision would be extremely challenging.

Other 3D coordinate systems have been developed for different applications. For example, *spherical coordinates*, with axes representing latitude, longitude, and elevation, conveniently describe location on the surface of the Earth. A simple path in spherical coordinates, such as "1 km east", is much more difficult to communicate using Cartesian coordinates. Similarly, coordinate systems for the human body should ideally have axes that follow anatomical structures, genetic expression patterns, chemical gradients, and/or other biologically relevant pathways. Several organ-specific coordinate systems have been developed, such as Talairach coordinates for functional brain imaging, but these do not extend to the entire body [Talairach 1967].

## 2   Vascular based coordinate system

In 2017, a Common Coordinate Framework (CCF) Meeting was organized by the National Institutes of Health, the Broad Institute, the Sanger Institute, and the Chan Zuckerberg Initiative. Meeting participants, which included anatomists, pathologists, clinicians, and technology experts from around the world, suggested several approaches for mapping the cells of the human body, including anatomically based coordinate systems [CCF Meeting 2017]. One of these, which uses the vasculature, was further discussed at a 2019 CCF Workshop, specifically for localizing kidney cells [CCFWS-01 2019]. In this manuscript, we present a conceptual overview of this vascular based coordinate system and describe its benefits and limitations (**Figure 1**).

The goal of a vascular coordinate system is not to build a map of the vasculature. Rather, the idea is to use known vascular pathways through the body as an axis in a coordinate system that can describe the position of anatomical elements, such as cells, in the *tissue surrounding the vasculature*. In this proposed framework, the heart may be viewed as the origin in the whole-body map. The vascular axis contains vascular loops (comparable to "roads" in the Google Map metaphor) that extend out through the aorta or pulmonary artery, gradually taper off to single cell size as they approach every other cell in every organ in the body, and return to the heart. Along these pathways, the vasculature follows the unique biology of the various organs and tissues; and, likewise, the organs and tissues of the body could not exist without the unique properties of their closely integrated vasculature.

Just as the genome map unfolds the DNA into a linear sequence, we can imagine unfolding the complex 3D twists and turns of the vasculature axis into a simpler 2D "hub-and-spoke" shape with the heart in the center. This schematic representation of the vasculature makes it easier to (1) describe location in the body; (2) align vascular pathways with varying 3D shapes but equivalent function from different people; (3) identify patterns such as changes in cell type and gene expression as one transitions from larger to smaller vessels within an organ (along a spoke); and, (4) compare tissue at the same level (e.g., capillaries) in different organs (across spokes).

An organ, such as the kidney, has tens of thousands of vasculature pathways if every capillary is considered distinctly. However, in order to make the construction of a CCF feasible, we will make an additional simplifying assumption that many of these pathways are indistinguishable from each other. Notably, the smallest vessels (capillaries) are at the center of the small specialized functional units in each organ (e.g., hepatic lobules in the liver, glomeruli in the kidney, and alveoli in the lung). A whole organ requires many of these vascular-centered functional units that are anatomically and physiologically similar [de Bono 2013]. This similarity can be used to collapse numerous vascular pathways into a few representative ones (**Figure 2**). In the future, as we learn more about the biomolecular profiles of these pathways, we can refine the CCF by splitting the pathways into any subtypes that are discovered.

An "address" of an individual cell in a vascular coordinate system has four components. (1) The first is the nearest anatomically named vessel that most precisely identifies a vascular loop (the vascular territory or "watershed"), such as the right renal artery. (2) The next component represents the branching level of the vessel. In the kidney, examples from proximal to distal include the interlobar arteries, cortical radiate arteries, afferent arterioles, and glomerular capillaries. The branching level can also be defined numerically (e.g., Strahler number), with capillaries at the zero level, and higher levels with larger numbers. (3) The third component is the perpendicular distance in micrometers or number of cells from the endothelial layer of the vessel. (4) A fourth component could be the rotational angle perpendicular to the length of the vessel. Zero degrees can be defined as the direction towards an anatomical structure, such as the center of the nearest glomerulus in the kidney. The utility and need for these components would need to be confirmed by experimental testing.

Another approach is to use endothelial cells as "anchoring cells" to identify the relative positions of the other cells in a GPS-like fashion. Endothelium lines the entire vasculature. Distinct endothelial cell (EC) types appear in different parts of the vasculature, which can be detected histologically through unique cell characteristics and neighboring cells (the "histologic fingerprint"), as well as through analysis of their specific gene profiles (the "gene expression fingerprint") [Guo 2012, Vanlandewijck 2018, Jourde-Chiche 2019]. Vascular organizational pattern, branching complexity, diameter, etc. provide additional recognizable features that form a "vascular architecture fingerprint" for each tissue [Johnson 2003, Boseetti 2016]. Each tissue has specific oxygen gradients, which can be used to determine distance from the vessel. When PO2 cannot be measured directly, one can infer the cell distance from the induction/expression gradients of hypoxia-inducible genes in various types of cells (the "hypoxia fingerprint") [De Santis 2015, Patnaik 2009, Koch 2002].

## 3    Benefits of a vascular CCF

A vasculature coordinate system makes sense biologically. Every living cell must be within a small radius of the closest blood vessel (100μm to 1mm, depending on the tissue) in order to receive oxygen [Tsai 2003]; and, every vessel is lined up by the same continuous layer of endothelium [Hacking 2019]. Thus, the vasculature forms an unbroken pathway that reaches all parts of the body, seamlessly tapering down across scales, from macroscale (whole body/clinical) to mesoscale to microscale (single cell-size capillary vessels). As part of this, it handles the gradual transition from large arteries and veins, which are conserved in most people, down to the millions of microscopic vessels that are recognizable by category (e.g., glomerulus capillary) but not by individual names. The vascular system adapts to individual body size and shape, accounting for inter-individual variations [Paruchuri 2015, Abadi 2018]. It defines the shape of functional units in different organs, such as liver lobules, kidney glomeruli, and lung alveoli. Normal development of new tissue begins with vessel formation before additional cell types can grow around it [Si-Tayeb 2010].

The vascular system also reaches all the organs, tissues, and cells in a contiguous fashion. Other anatomical structures do not seem to have similar versatility. For example, the skeletal system extends to the entire body; however, the distance from a cell within an organ to the nearest bone can be several centimeters, and that bone is unlikely to be part of the organ, and even less likely to be part of a tissue specimen. The nervous system follows many of the same pathways as larger vessels invested with contractile properties, but it is not known whether they extend to follow the smaller vessels that reach all tissue cells. Thus, unlike cells' dependence on oxygen, there isn't a guarantee that every cell is in close proximity to a nerve. Organ-specific reference systems are based on unique structural characteristics that exist only in that individual organ (e.g., specific regions of the brain) and are unlikely to be useful for representing cell locations elsewhere in the body.

There are many practical benefits of a vascular coordinate system. Vascular pathways in all organs have been studied extensively and are described in the literature in detail. They are useful, well known, and have standardized names across many specialty domains. Tissue vascularization patterns are used clinically to diagnose disease; and, vessels are used in surgery and biopsies as the main anatomical landmarks and to define "vascular territories" [Taylor 1987, Majno 2014]. Although the exact position in 3D Cartesian space might be difficult to determine for a resected tissue specimen, the surgeon will typically be able to indicate which vascular watershed included the specimen.

## 4    Limitations of a vascular CCF

Only knowing the base pairs in a genome sequence does not provide enough information to determine how the DNA is folded in a particular cell. Similarly, although the unfolded vascular coordinate system can describe the 3D spatial relationship of nearby cells at the microscale, the winding paths that vessels take and the relative distance between cells at larger scales is lost. Information about larger scale structures could be obtained by integrating clinical imaging and histological and molecular patterning.

Some additional positional information is lost by collapsing vascular pathways. For example, a vascular coordinate system might initially treat all glomerulus capillaries in the kidney as the same structure, potentially masking differences in glomeruli at the superior and inferior renal poles. Though, as we learn more about the biomolecular profiles of these pathways, we can iteratively refine the vascular CCF over time by splitting pathways into any subtypes that are discovered.

Despite these limitations, a vascular coordinate system retains enough positional information to answer many types of research questions, by allowing researchers to precisely localize individual cells within functional units, tissues and organs, or compare the effects of context on various cell types across the whole body at the macro- and meso- scales.

## 5    Summary

The vascular system has several properties that respond to the characteristics previously deemed as desirable for a CCF to enable mapping all the cells in the human body [CCF Meeting 2017]. (1) It works across several scales. Through a continuous endothelium it makes a seamless transition from large anatomical structures down to the cellular level. (2) It is applicable to all body tissues. Because organs develop around vessels, the vasculature frames organ architecture at all scales. Vascular patterns are so distinctive that images of vessels alone, with all other cells removed, can easily be used to identify the organ. (3) It accounts for donor differences. Vascularization naturally adapts to individual variations in tissue, body size and shape. Instead of having to map the 3D coordinates of each glomerulus in one person to those in another person, the vascular CCF would position each cell within a representative glomerulus capillary from each person. (4) The vasculature is well known to practitioners of many different clinical specialties, helping both diagnose disease and guide surgeries, and used as landmarks by pathologists, radiologists, and other clinicians. This would facilitate positioning tissue samples in the body and registering the location of individual cells. While no single CCF is ideal for all use cases, a vascular CCF defines a natural coordinate system that would make it easy to combine biomolecular data from multiple people and ask biologically relevant research questions.

## 6 Conflict of Interest

The authors declare that the research was conducted in the absence of any commercial or financial relationships that could be construed as a potential conflict of interest.

## 7 Author Contributions

GW conceived this study based on reports from the 2017 CCF Meeting and 2019 CCF Workshop. KB and YJ provided critical feedback. GW wrote the initial draft, with additional text transcribed from the CCFWS-01 slide presentation. All authors contributed to editing the manuscript.

## 8 Funding

This research is supported by the NIH Common Fund, through the Office of Strategic Coordination/ Office of the NIH Director under award OT2OD026671.

## 9 Acknowledgments

We thank Dr. Zorina S. Galis from the National Institutes of Health for helping develop the idea of a vascular-based coordinate system at the 2017 CCF Meeting and 2019 CCF Workshop, providing us with presentations from these events, and participating in numerous discussions that formed the basis of this manuscript. Dr. Marc Charette provided comments on an earlier draft of the paper.

**Figure 1. The hub-and-spoke structure of a vascular coordinate system.** The vascular coordinate system consists of representative vessel loops that begin and end at the heart and extend to the functional units of each organ (counterclockwise from upper-left: alveoli, glomeruli, and hepatic lobules). The "address" of a cell includes of the name of the loop (e.g., "R" for renal, "C" for coronary) and the branching level of the nearest vessel (e.g., "RA1" for afferent arterioles in the renal loop). There are multiple ways to pinpoint an even more precise "GPS-like" position, including, going clockwise from upper-right, the "hypoxia fingerprint" that results from decreasing oxygen levels farther from the vessel, the unique "histologic fingerprint" and "gene expression fingerprint" in different types of vascular endothelial cells, and the distinctive "vascular architecture fingerprint" found in different types of tissue. The vasculature extends to all parts of the body (bottom) and frames all organs at all scales.

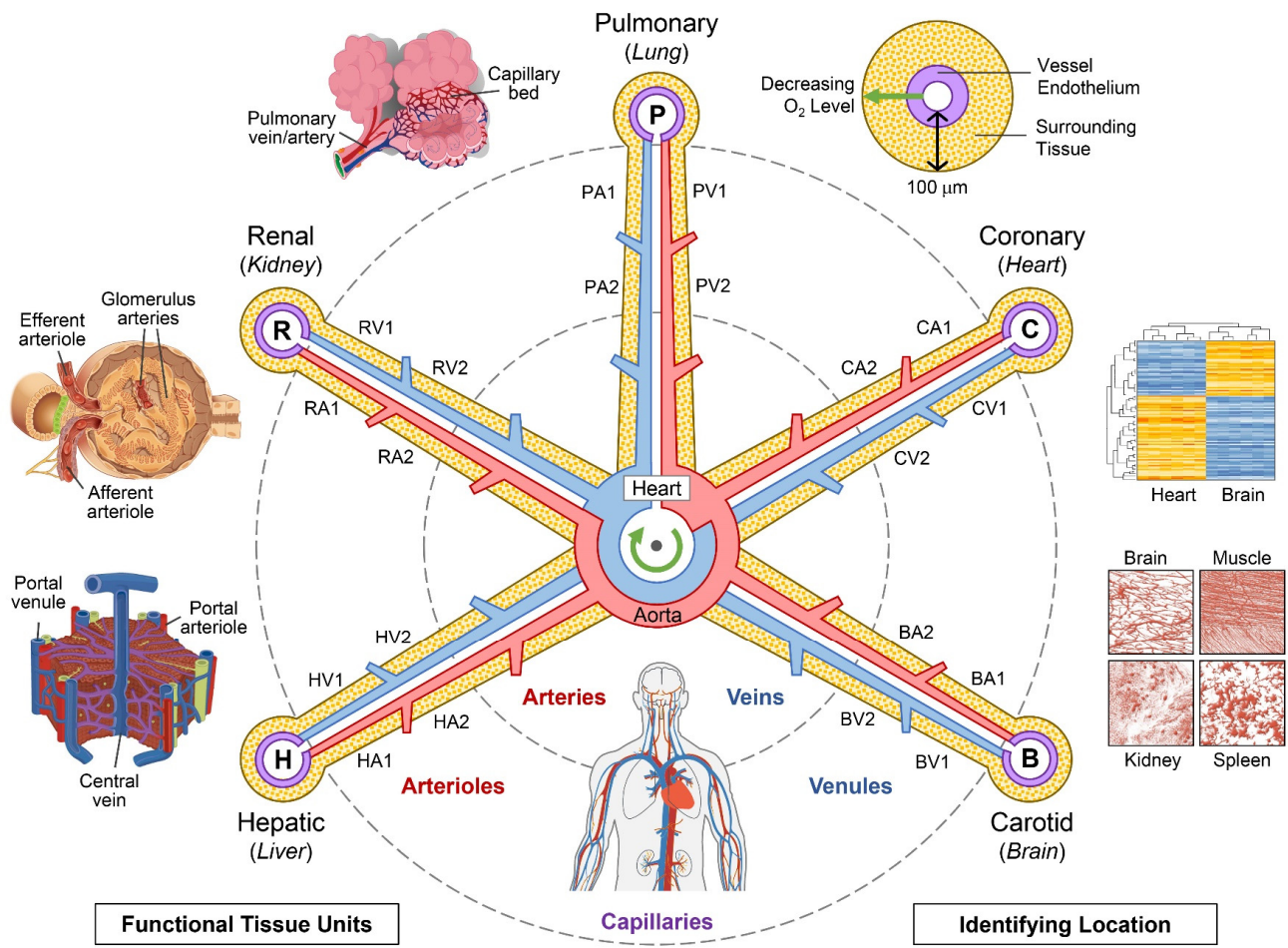

Source Files

Images of the hepatic lobules, glomeruli, alveoli, gene expression heat map, and whole body vasculature are adapted from Wikimedia Commons: https://commons.wikimedia.org/wiki/File:2423_Microscopic_Anatomy_of_Liver.jpg (Public Domain); https://commons.wikimedia.org/wiki/File:Juxtaglomerular_Apparatus_and_Glomerulus.jpg (CC BY 3.0 license); https://commons.wikimedia.org/wiki/File:Alveolus_diagram.svg (Public Domain); https://commons.wikimedia.org/wiki/File:Iris_dendrogram.png (CC BY-SA 4.0 license); and, https://commons.wikimedia.org/wiki/File:Circulatory_System_en.svg (Public Domain). Gene expression heat map recreated from Guo 2012 using simulated data. Vascular architecture based on Bosetti 2016 (CC BY 4.0 license).

**Figure 2. Unfolding the renal vascular pathway down to the single cell level.** A representative vascular pathway enters the kidney through the renal artery, passes through the glomerulus, and returns out the renal vein. At the macroscale (left) and mesoscale (center), labels in yellow boxes are anatomical structures that correspond to different vessels along this loop. At the microscale (right), labels in yellow boxes are different types of cells that are within a short distance of a nearby vessel in the glomerulus. The position of individual cells can be described by the nearest vessel's address (e.g., "RA0" for glomerulus capillary), the distance in micrometers or number of cells from the vessel endothelium and an angle perpendicular to the length of the vessel (e.g., "RA0-10µm-135°" or "RA0-1c-135°"). On the far right, each vessel and its surrounding tissue has been extracted from the original image and aligned vertically to make it easy to see how cell types and distribution change along the vascular pathway. In this example, the aligned images have also been rotated so that the center of the glomerulus is always on the right.

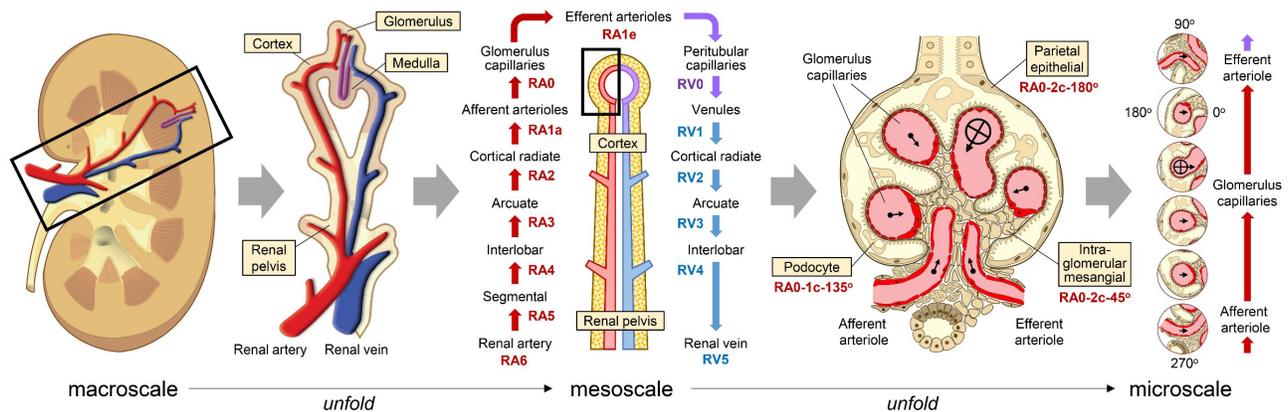

Source Files

Images of the kidney and glomerulus are adapted from Wikimedia Commons:
https://commons.wikimedia.org/wiki/File:KidneyStructures_PioM.svg (CC BY 3.0 license); and,
https://commons.wikimedia.org/wiki/File:Renal_corpuscle-en.svg (CC BY-SA 4.0 license).